\documentclass{vgtc}                          % final (conference style)
\graphicspath{{figures/}{pictures/}{images/}{./}} % where to search for the images

\usepackage{times}                     % we use Times as the main font
\usepackage{tabu}                      % only used for the table example
\usepackage{booktabs}                  % only used for the table example
\usepackage{lipsum}                    % used to generate placeholder text
\usepackage{mwe}                       % used to generate placeholder figures

\usepackage{mathptmx}                  % use matching math font
\usepackage{xcolor}
\usepackage{caption}
\usepackage{subcaption}
\onlineid{1596}

\vgtccategory{Research}

\vgtcinsertpkg

\title{Secure Text Entry using a Virtual Radial Keyboard with Dynamically Resized Keys and Non-Intrusive Randomization}

\author{Yuxuan Huang\thanks{e-mail: \{huan2076, nie00035, interran, suma\}@umn.edu} 
					\and Qiao Jin\thanks{e-mail: qjin4@ncsu.edu}
					\and Tongyu Nie$^{*}$ 
					\and Victoria Interrante$^{*}$ %\hspace{.75in}
					\and Evan Suma Rosenberg$^{*}$
			 }
\affiliation{\scriptsize  $^{*}$University of Minnesota \hspace{.4in} $^{\dag}$North Carolina State University \hspace{.4in} }

\teaser{
  \centering
  \includegraphics[width=\linewidth]{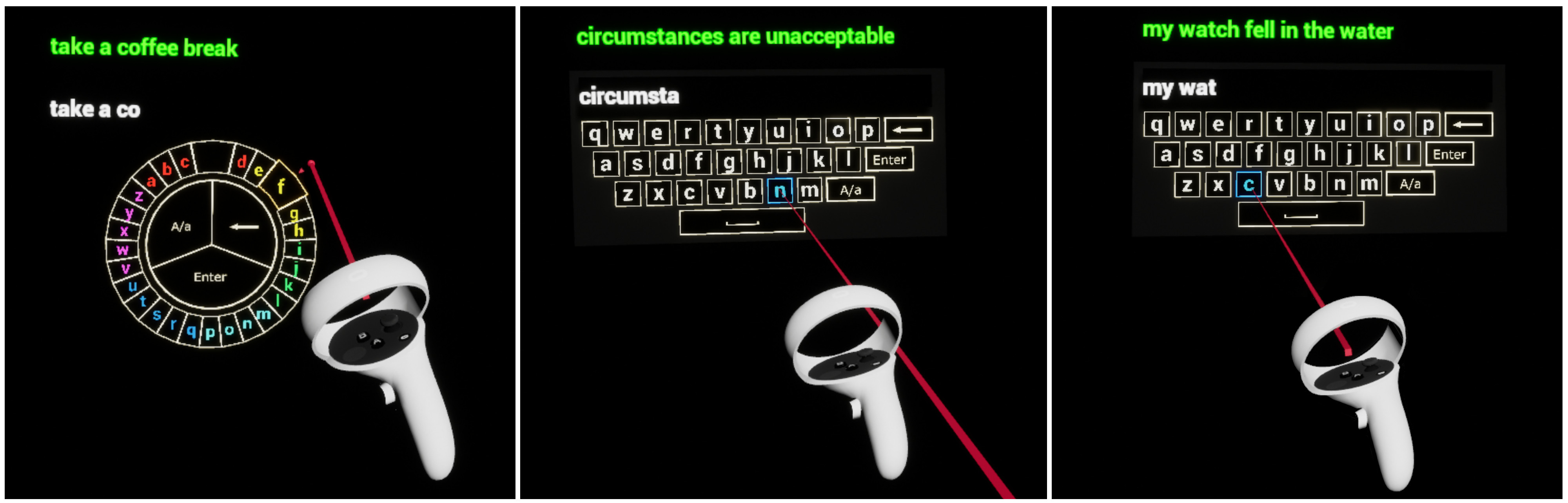}
  \caption{The radial keyboard proposed in this paper (left) vs. a state-of-the-art secure VR text entry method ``Intermittent Starting Point Randomization''~\cite{wan2024design} (middle) vs. standard QWERTY keyboard (right).}
  \label{fig:teaser}
}

\abstract{
  As virtual reality (VR) becomes more widely adopted, secure and efficient text entry is an increasingly critical need.
  In this paper, we identify a vulnerability in a state-of-the-art secure VR text entry method and introduce a novel virtual radial keyboard designed to achieve a balance between security with usability.
  Keys are arranged alphabetically in a circular layout, with each key selected by controller rotation and dynamically expanding to facilitate precise selection. 
  A randomized rotation mechanism shifts the keyboard after each keystroke, preserving relative key positions while disrupting absolute spatial mappings to protect against inference attacks.
  We conducted a within-subject study (N=30) comparing our method with the prior secure technique and a standard QWERTY keyboard. 
  Results showed that the radial keyboard significantly improves resistance to keystroke prediction attacks while incurring a tradeoff in entry speed and subjective workload due to the unfamiliar non-QWERTY layout.
  However, both quantitative trends and qualitative feedback indicate strong potential for performance improvements with practice. 
  We also discuss design implications, possible interface refinements, and directions for future work, including layout variations and visual enhancements.
} % end of abstract


\begin{document}

%% The ``\maketitle'' command must be the first command after the
%% ``\begin{document}'' command. It prepares and prints the title block.

%% the only exception to this rule is the \firstsection command
\firstsection{Introduction}

\maketitle

% Introduce the importance of VR text entry
With Virtual Reality (VR) technology becoming more accessible and widely adopted in various fields~\cite{hamad2022virtual}, VR has demonstrated great potential in playing a more important role in our work and life.
As VR applications expand beyond entertainment into areas like office work~\cite{grubert2019office}, collaboration~\cite{jin2023collaborative}, training and education~\cite{jin2022will}, the need to support essential tasks such as text entry is becoming increasingly critical~\cite{dube2019text}. However, significant challenges remain in terms of both usability and security for VR text entry~\cite{knierim2020opportunities,wan2024design}.

% Usability Challenges (Entry efficiency, error rate, user experience)

%Most existing VR applications rely on virtual keyboards for text entry, but research indicates that typing with virtual keyboards is significantly slower, more error-prone, and more fatiguing compared to physical keyboards in traditional, non-immersive environments~\cite{dube2023ultrasonic}.
%The primary disadvantage of virtual keyboards stems from the absence of haptic feedback. With physical keyboards, experienced users rely heavily on the tactile sensation of physical keys to differentiate between adjacent keys and type efficiently. However, this tactile familiarity does not translate to virtual keyboards, where users must depend solely on visual cues~\cite{chen2019exploring}.
Text entry in VR is inherently less efficient compared to using physical keyboards or touchpads. The root cause lies in the difficulty in precisely selecting the keys in an efficient manner as with the physical keyboard~\cite{dube2019text} due to the lack of haptic feedback and compromised eye-hand coordination~\cite{knierim2018physical,grubert2018effects}. Consequently, many methods incorporate haptic feedback by using physical keyboards~\cite{grubert2018text,knierim2018physical,otte2019towards,pham2019hawkey} or other haptic devices~\cite{dube2023ultrasonic,gupta2020investigating} to enhance the differentiation between keys and improve selection accuracy. However, the dependency on additional hardware such as a physical keyboard restrict their usage to specific scenarios, such as when the user is seated at a desk where the keyboard can be placed, limiting their practicality in non-stationary VR experiences~\cite{speicher2018selection}. Therefore, most existing VR applications rely on virtual keyboards for text entry, and optimizing the usability of virtual keyboard-based entry has been an important topic~\cite{jiang2020hipad,leng2022efficient,wang2023eye,liu2024crosskeys}.

% Security Challenges
In addition to usability, the security aspect of text entry in VR has also attracted increased attention in recent years. While passwords remain the most common form of sensitive input, other types of sensitive information, such as personal details, medical records, and confidential data from corporations or government agencies, are frequently entered via natural language~\cite{wan2024design}. Compromise of such natural language text entry can lead to serious consequences, ranging from scams and financial losses to threats to national security~\cite{wan2024design}, and a secure text entry method is a prerequisite for VR to be adopted in such contexts.
Research has shown that the deterministic spatial relationships of the keys can be exploited, allowing effective prediction of the keyboard layouts and keystrokes. This issue is particularly pronounced in VR due to slower text entry rates, larger required body movements, reduced environmental awareness, and the presence of built-in sensor tracking, distinguishing it from traditional text entry scenarios such as those on smartphones or physical keyboards. 
For example, Ling et al.\cite{ling2019know} proposed a keylogging inference attack that utilizes the orientation data of the headset and controller during text entry. These orientations can be inferred from recorded videos using computer vision algorithms or directly extracted from the controller's sensor data. Similarly, Wu et al.\cite{wu2023privacy} introduced a two-step keystroke snooping method. This method first reconstructs the virtual keyboard using onboard sensor data and then predicts the entered content. Their experiments demonstrated an 80–90\% accuracy in recovering text or passwords. %Alarmingly, they also discovered that access to sensor data does not require user permission in mainstream VR software development kits such as OpenVR, Oculus, and WebXR, creating significant opportunities for such attacks.
Similar keystroke inference attacks in VR have been widely discussed in other work, including~\cite{al2021vr,meteriz2022keylogging,luo2022holologger,gopal2023hidden,lee2023vrkeylogger, zhang2025airtypelogger}, etc. To counteract this vulnerability, research has focused on introducing randomization to disrupt these predictable spatial patterns, but naively randomizing keyboard layouts often leads to poor usability, as users struggle with unfamiliar and inconsistent designs~\cite{mackenzie2001empirical}. Consequently, the primary challenge in addressing the security problem lies in implementing randomization strategies that minimize negative impacts on usability, but existing work that specifically considers the security challenge in VR text entry is limited~\cite{wan2024design,wan2024analysis}.

%A common factor among these attacks is their reliance on the fixed layout and location of virtual keyboards, which creates a strong spatial correlation between the pose of the user or entry device and the key being selected. As highlighted in~\cite{meteriz2022keylogging}, potential countermeasures include randomizing the keyboard layout or location to disrupt the deterministic mapping between 3D positions and specific keys. However, such randomization can negatively impact usability and typing performance~\cite{mackenzie2001empirical}. Balancing usability, performance, and security to design a text entry method for VR that excels in all three aspects remains a complex and challenging problem. 

% 

%Others adopt virtual keyboards and introduce alternative interaction techniques that are less error-prone than conventional point-and-click methods, such as using wrist rotation~\cite{liu2024crosskeys} or the controller touchpad~\cite{jiang2020hipad}, or they employ custom keyboard layouts designed to reduce the likelihood of mis-entry~\cite{leng2022efficient,wang2023eye}. These approaches leverage the unique affordances of VR, where the interactions and virtual keyboards are not bound by physical constraints and can be optimized for spatial manipulation. Since the spatial interactions in VR differ fundamentally from traditional user interfaces, conventional designs may not translate effectively. Exploring VR-specific spatial manipulations that enhance selection precision remains a promising direction for advancing VR text entry.

In this paper, we propose a novel radial keyboard to address the two challenges discussed above, i.e., improving precise selection and enhancing security of text input through non-intrusive randomization. The specific contributions of this paper include:
\begin{enumerate}
    \item We present a keystroke prediction strategy to illustrate the vulnerability of natural language text entry with limited randomization~\cite{wan2024design,wan2024analysis} and demonstrate its effectiveness against the state-of-the-art~\cite{wan2024design}.
    \item We propose a secure virtual radial keyboard that is not only robust against keystroke snooping attack but also relaxes the precision requirement of key selection by resizing the keys in a random but non-intrusive fashion.
    \item We conducted a user study to compare the proposed radial keyboard with the state-of-the-art~\cite{wan2024design}, and standard QWERTY keyboard in terms of both usability and security.
\end{enumerate}

\section{Related Work}

\subsection{Typing Methods for Virtual Keyboards}
% Different Interaction Methods

Various typing methods have been explored for virtual keyboard input. According to~\cite{wan2024analysis}, mainstream typing techniques in VR can be categorized into position-based and ray-based approaches.

% Position-based
With position-based text entry, users interact directly with the virtual keyboard by tapping or swiping on the keys using either their controllers~\cite{chen2019exploring, spiess2022direct, boletsis2019controller, boletsis2019text, dube2019text, yao2020punch, akhoroz2024poke} or their hands~\cite{gupta2020investigating, wang2023comparative, dube2020impact, akhoroz2024poke}. This method feels natural and intuitive, as it closely mimics how people interact with physical objects in the real world.
However, position-based text entry tends to be relatively slow and can cause fatigue due to the extensive movements required, particularly when the keyboard is positioned mid-air~\cite{speicher2018selection}.

% Ray-based
In this paper, we focus on ray-based text entry using controllers. With ray-based text entry, users select keys by pointing at them with a virtual ray, typically controlled by a controller~\cite{chen2019exploring, wan2024analysis, wan2024design}, head movements~\cite{yu2017tap, xu2019ringtext, lu2020exploration}, or gaze direction~\cite{rajanna2018gaze, he2022tapgazer}. Compared to position-based methods, ray-based approaches require significantly less physical movement, as adjusting the ray's direction involves only small rotations of the wrist, head, or eyes. This reduction in movement not only minimizes fatigue but also enhances security against direct observation from bystanders, as the subtle movements are harder to detect and exploit.
% Each ray-based approach has its own advantages and limitations:
% \begin{enumerate}
%     \item Controller-based ray-casting can be efficient and dexterous, as small wrist rotations result in significant directional changes in the ray. However, this same characteristic makes precise selection challenging, and the small angular size of keys further exacerbates the difficulty~\cite{kopper2010human}.
%     \item Head-directed selection frees the hands for other tasks, but is less efficient and intuitive since humans are not naturally accustomed to using neck rotation for precise targeting, which can lead to neck strain or motion sickness with prolonged use~\cite{blattgerste2018advantages, xu2022evaluation}.
%     \item Gaze-directed selection is highly intuitive and user-friendly but requires eye-tracking technology, which increases cost and may not be available on all VR devices.
% \end{enumerate}

%The design of a text-entry method comprises two key components: the user interface and the user interaction. The user interface pertains to the arrangement and presentation of virtual keys, while the user interaction involves the mechanisms by which keys are selected and entered.

\subsection{Alternative Virtual Keyboard Layouts}
%\subsection{Methods to improve selection accuracy}
% QWERTY
The virtual keyboard in most existing VR applications are using the QWERTY layout~\cite{dube2019text,speicher2018selection}, which has been the standard layout for physical keyboards for decades~\cite{noyes1983qwerty}. The apparent advantage of adopting it in VR is the users' familiarity, which enhances efficiency and user acceptance~\cite{dube2019text}.
However, using the virtual keyboard in VR is a completely different experience from using physical keyboards. With the QWERTY layout, the keys are usually small and closely-spaced. When typing with a physical keyboard, the haptic feedback helps differentiate between keys, and the dexterity of the fingers allows for freehand ``ten-fingered'' entry, which together reduce error rates and improves efficiency. In contrast, when typing on a similarly designed virtual keyboard, the absence of haptic feedback, the lack of direct view of the hands, the latency and limited reliability of hand-tracking system, or the difficulty in aiming precisely at small targets with a ray make accurate and efficient typing much more challenging and error-prone~\cite{dudley2023evaluating}.
%Furthermore, most keylogging inference attacks target keyboards with QWERTY layouts~\cite{wu2023privacy,meteriz2022keylogging}, posing a significant security risk to virtual keyboards in VR environments.

% alternative (radial, flower, cubic)
Consequently, some studies have moved away from the standard QWERTY layout to explore alternative designs for virtual keyboards tailored to the unique advantages and limitations of VR.
For instance, to address the difficulty of precise selection in VR, some researchers have proposed custom keyboard layouts that maintain a strong resemblance to QWERTY but employ different spatial arrangements. Examples include layouts in a flower shape~\cite{leng2022efficient} or an eye shape~\cite{wang2023eye}, where keys are spaced farther apart to reduce errors and improve usability.

Some other studies abandon the QWERTY layout entirely in favor of radial designs, where keys are arranged alphabetically and differentiated by 1D angular differences instead of 2D positional differences to relax the precision requirement of aiming at a key.
For example, Yu et al. introduced ``PizzaText''~\cite{yu2018pizzatext}, in which letters are grouped into sets of four and distributed across seven sectors. Within each sector, the four letters are mapped to up, down, left, and right, respectively. Users input letters by pushing one joystick to select the sector, and using the other to specify the letter within that sector.
Similarly, Nguyen et al. proposed two text entry methods utilizing radial keyboards~\cite{nguyen2020text}. In these approaches, the radial keyboard is mapped onto an HTC Vive controller touchpad. To input a letter, users first tap the touchpad to select the sector containing the letter. They can then either choose the letter from a secondary menu that appears, or tap the sector repeatedly—$n$ times for the $n$-th letter in the sector.
Jiang et al.~\cite{jiang2020hipad} extended this concept with word prediction feature to further increase entry efficiency.

In addition to 2D virtual keyboards, some studies have explored potential 3D layouts.
Liu et al. introduced ``CrossKeys''~\cite{liu2024crosskeys}, where letters and symbols are grouped similarly to the approach in ``PizzaText''~\cite{yu2018pizzatext}, but are arranged along two intersecting arc segments. These groups are mapped to horizontal (yaw) and vertical (pitch) rotations of the controller.
Yanagihara and Buntaru proposed a 3D cubic keyboard prototype in which keys are arranged within a $3 \times 3 \times 3$ cube. Text entry is facilitated through a combination of controller dwelling and button interactions~\cite{yanagihara2018cubic}. However, formal evaluations are required to better understand the feasibility and practical applicability of such a layout.

%commented out by ESR to save space
%The radial keyboard we propose is inspired by these work where the keyboard layouts and interactions are optimized for spatial manipulation in VR, as it differs fundamentally from interacting with traditional user interfaces. 
%Exploring VR-specific spatial manipulations that enhance selection precision remains a promising direction for advancing VR text entry.

\subsection{Security in VR Text Entry}
Besides usability, a few recent works also considered the security aspect of text entry in VR. 
As mentioned earlier, these interfaces are vulnerable to keystroke inference attacks~\cite{meteriz2022keylogging,luo2022holologger,gopal2023hidden,lee2023vrkeylogger}. Sensitive information can be exposed to malicious attackers through controller tracking data~\cite{ling2019know,wu2023privacy}, video recordings~\cite{ling2019know,zhang2025airtypelogger}, and other easily obtainable information~\cite{al2021vr}.

% Randomization / Redirection
To counter these keystroke prediction methods, randomization strategies have been proposed to break the deterministic spatial relationships between the gestures and keystrokes, while minimizing intrusiveness.
For example, in \cite{wan2024design}, Wan et al. proposed different randomization strategies for ray-casting-based keyboard entry, such as randomizing the direction or the starting point of the ray.
Similarly, Wan et al. also proposed randomization strategies for the properties of the virtual keyboard, such as the location of the keyboard or the spacing between the keys~\cite{wan2024analysis}, while maintaining the QWERTY layout.
Since the randomization outcomes are unknown to anyone other than the VR user, existing keystroke prediction methods fall short because they are unable to account for the random factor during text entry to predict which key is being pointed at. 
To reduce intrusiveness, the randomization is introduced at random intervals instead of after every key press, which allows for a more coherent typing experience while preventing attackers from knowing when the randomization happens.

These randomization methods are inspirational, but a major limitation is that they can only support a limited amount of randomization due to usability constraints. Despite being secure against basic prediction methods not accounting for the randomization, we found that the small randomization space can be effectively compromised by uniform sampling. The specific sampling attack method to compromise the security mechanism will be discuss in section \ref{sec::attack}.
Therefore, in light of the absence of a truly secure text entry method in VR, in this paper we propose a novel text entry method that supports full randomization of the spatial location of the keys, while minimizing the intrusiveness of such randomization.

%Despite not targeting text entry on full keyboard, Weiss et al.~\cite{weiss2024exploring} proposed using hand redirection for PIN entry, where the user's virtual hand representation is redirected away from the actual hand location with a random offset when they reach out to select a key, so that they will be reaching to different random locations even when selecting the same key, thus improving its robustness against observation attack..

% ==================== Uniform Sampling Attack ======================
\section{Uniform Sampling Attack}
\label{sec::attack}
In this section, we introduce the uniform sampling attack scheme to demonstrate that while existing methods are effective for password-style inputs~\cite{wan2024analysis,wan2024design}, they still face challenges when handling general text-based input. Specifically, we use the method that achieves best overall performance in terms of security and usability in~\cite{wan2024design}, i.e., the Intermittent Starting Point Randomization (ISPR), as example, but the same strategy can be applied to other methods featuring intermittent randomization in a limited range as well.

ISPR is a ray-based text entry method using a virtual QWERTY keyboard, where the starting point (SP) of the ray jumps forward or backward within a certain range after typing a random number of letters. With ISPR, even though the direction of the ray is exposed by the controller rotation, the starting point of the ray is unknown, so existing prediction algorithms that assume the ray starting from the controller will produce incorrect results.
In~\cite{wan2024design}, the keyboard is positioned 10 meters in front of the user, with each letter key being $0.3m\times0.3m$. The SP randomization range is $[-5m, 5m]$ along the global forward axis, and the randomization is triggered after a random number of key presses drawn from the range [4,12].
According to \cite{wan2024design} and \cite{wan2024analysis}, ISPR and the other secure entry methods based on intermittent randomization are not suitable for password entry due to the infrequent randomization. The uniform sampling attack is also intended for natural language reconstruction rather than predicting character sequences without a clear meaning.

The idea of the uniform sampling attack is that even though the exact SP location is unknown, it can be approximated by a sample close to it. By uniformly sampling the entire randomization space, there will always be a sample close enough to the actual SP. For example, with a random range of [-5,5], we can take 5 samples, i.e., {-4, -2, 0, 2, 4} to have a reasonable coverage of the possible SP locations.
Due to usability constraints, the randomization space in this type of method is intrinsically limited, so a small number of sample suffices. Although each prediction is likely only partially correct, combining them together may allow us to piece together the meaningful parts and recover the original sentence.

To examine the feasibility of the Uniform Sampling Attack, we conducted a small pilot study in which a single user entered 13 phrases using ISPR. The attacker reconstructed the keyboard by performing triangulation with multiple rays to measure the scale and distance, and approximated the keyboard orientation using the user's mean headset orientation during text entry. For each phrase, five uniformly sampled predictions were generated as described above, and the predictions were combined using ChatGPT with the following prompt: \textit{``You will be given 5 inaccurate predictions of the a sentence. Combine the meaningful parts to predict what the original sentence is. When confused, consider replacing letters with nearby letters on a QWERTY keyboard.''}
The attack successfully reconstructed 6 phrases verbatim, and several others with large portions correct. For example:
\begin{enumerate}
    \item mt vqnk azzoybr ia osqqqeaqn
    \item mt vank axxoybt ia odqqqrawn
    \item ny banj acckhbt us icqwqrsenl
    \item ng bsnj zcckhbg jd kcwewfxdn
    \item nh bxnn xvvjhbg jc jverdgcfnm
\end{enumerate}
combined into \textit{``my bank account is overdrawn''}.
A more formal testing of ISPR's robustness against the uniform sampling attack will be discussed and reported in section \ref{sec::study} and \ref{sec::res}, but this vulnerability we identified suggests that intermittent randomization within a limited range may be insufficient for security purposes. Therefore, we aim to enhance the security of VR text entry by introducing a more thorough randomization while keeping usability compromises at minimum.

% ======================= Radial Keyboard ===========================

\section{Radial Keyboard}

% \begin{figure}[ht]
%   \centering
%   \includegraphics[width=0.7\linewidth]{figures/RadialKeyboardIllustration.png}
%   \caption{The radial keyboard}
%   \label{fig:radial-keyboard}
% \end{figure}

We propose a virtual radial keyboard, where the letter keys are arranged in a circular, ring-like fashion in alphabetical order. The letter keys are divided into 6 groups of 4-5 letters. Each group of letters is visualized in a different color.
The space key is inserted among the letter keys in a random location and takes up twice the width of the letter keys to allow for more efficient identification and selection. Other commonly used non-letter keys, such as \textit{Enter} and \textit{Backspace} are positioned at the center of the circle. For the purpose of this study we did not include number or symbol keys, but extensions can be easily achieved by adding an additional key in the center to toggle between letters and numbers or symbols.
This radial layout offers two key benefits: 
\begin{enumerate}
    \item Intuitive Arrangement: Placing the letters in alphabetical order makes the layout more intuitive compared to non-QWERTY designs, reducing the learning curve for users.
    \item Usability-Preserving Randomization: Randomizing the rotation of the radial keyboard alters the location of the keys while preserving their relative order. This approach disrupts deterministic spatial patterns for enhanced security while minimizing usability drawbacks.
\end{enumerate}

% Entry
Users select the keys using ray-casting and enter them using the trigger button on the controllers.
For the keys positioned at the center, users point the ray directly at the key to select it. 
For the letter keys and the space, the selection area for each key extends infinitely outward from the circular keyboard. A letter key is considered ``under selection" if the intersection of the ray and the keyboard plane falls within the key's corresponding sector. The farther the ray intersects from the center, the greater the distance required to move to an adjacent key. This feature reduces the precision needed for key selection, offering a usability improvement not achievable with virtual QWERTY keyboards. See the left sub-figure of figure \ref{fig:teaser} for an illustration of key selection on the radial keyboard.

%The virtual radial keyboard incorporates two key mechanisms: dynamically resized keys and non-intrusive randomization.

\paragraph{Dynamically Resized Keys}
If a letter key is under selection, it dynamically expands to occupy an additional key width. This resizing enhances the user's visual perception of the selected key while further reducing the precision required for selection. The increased size also minimizes unintentional switching to adjacent keys caused by jittery hand movements or unsteady control. To avoid the expanded key from pushing the target key away from the cursor, the expansion will always be in the opposite direction.

\paragraph{Non-intrusive Randomization}

\begin{figure}[]
\centering
\begin{subfigure}{.32\columnwidth}
  \centering
  % include first image
  \includegraphics[width=\linewidth]{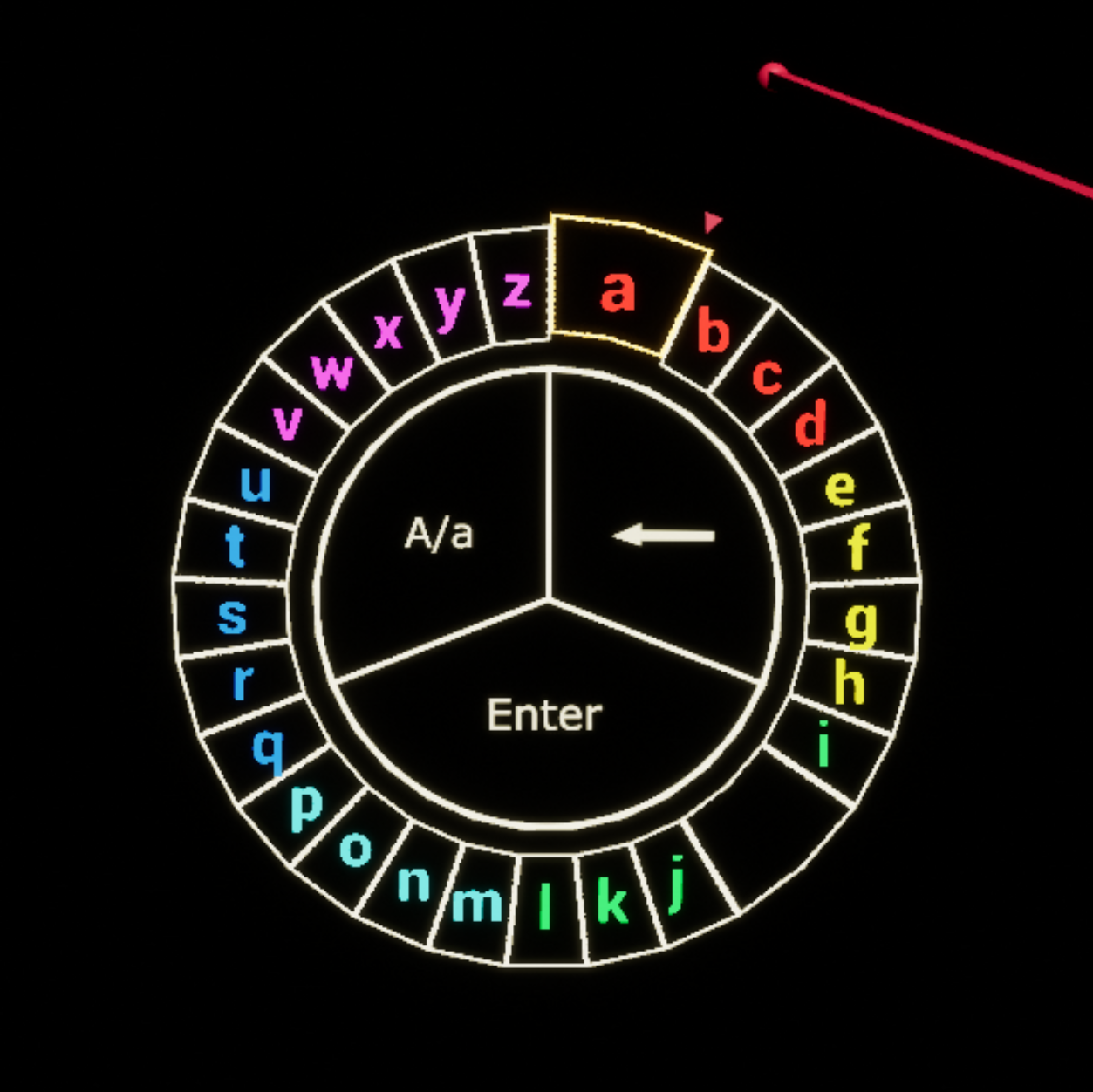}  
  \caption{user enters \textit{``a''}}
  \label{fig:random-a}
\end{subfigure}
\begin{subfigure}{.32\columnwidth}
  \centering
  % include second image
  \includegraphics[width=\linewidth]{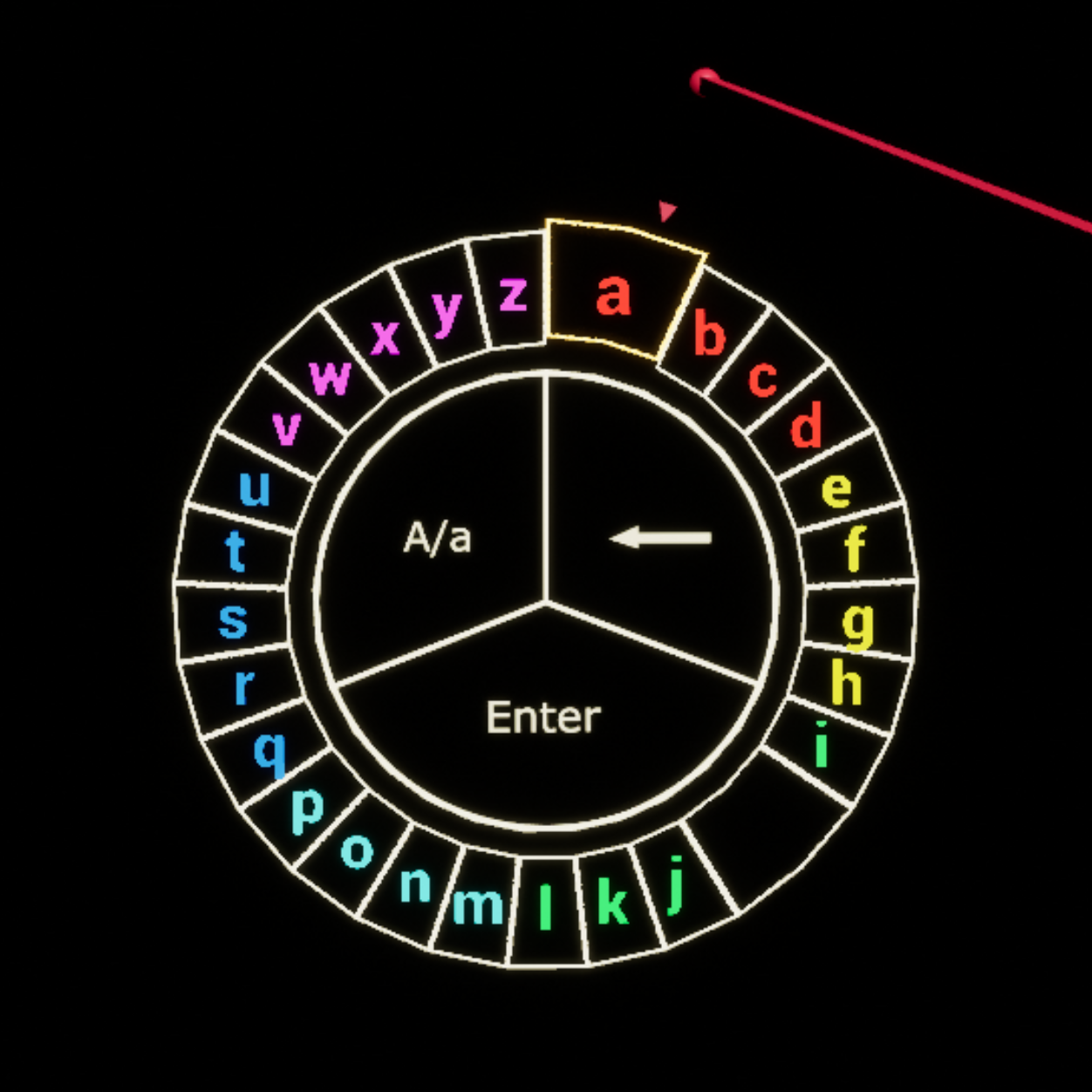}  
  \caption{possible outcome 1}
  \label{fig:random-b}
\end{subfigure}
\begin{subfigure}{.32\columnwidth}
  \centering
  % include fourth image
  \includegraphics[width=\linewidth]{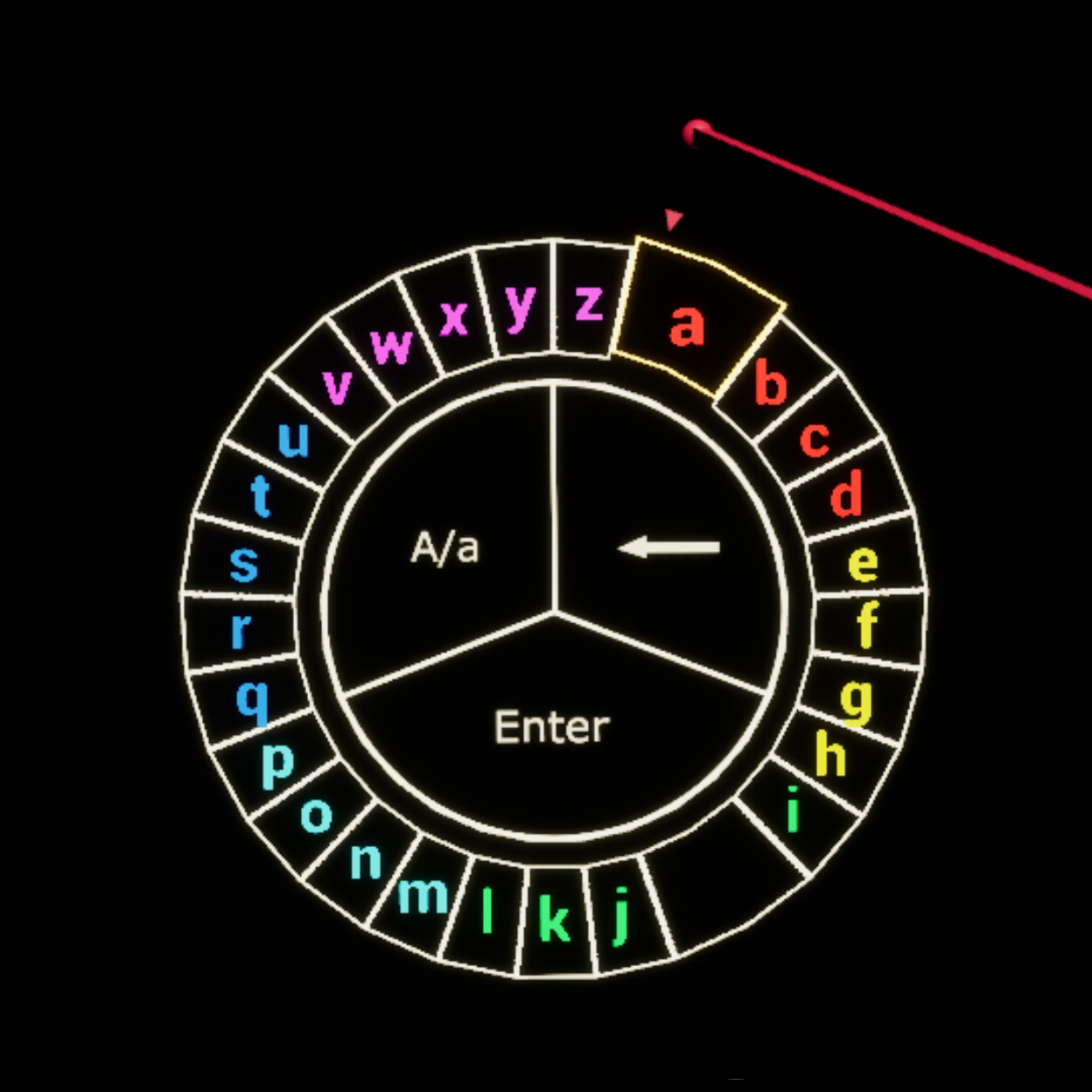}  
  \caption{possible outcome 2}
  \label{fig:random-c}
\end{subfigure}
\caption{(a) The user enters \textit{``a''} on the radial keyboard, which triggers the randomization and results in two possible keyboard configurations: (b) the selected key \textit{``a''} expands to the left of the cursor position, identical to the previous state (c) the selected key expands to the right of the cursor position, and all keys are shifted clockwise for one key width. }
\vspace{-8pt}
\label{fig:randomization}
\end{figure}

The location of the space key and the initial rotation of the keyboard are randomized, which makes the initial keyboard configuration unpredictable.
Moreover, whenever users enter a letter, randomization happens and the key under selection re-determines whether it expands to its left or to its right, relative to the location selected by the cursor (figure \ref{fig:randomization}).
Depending on whether the cursor is selecting the left or right half of the letter key, there will either be a shift to all the keys by one key width, or no change to the keyboard at all. The probability for the keys to expand towards the left over the right is randomly generated each time the keyboard is used in the range of [0.2, 0.8], so there can be a bias for the keys to rotate clockwise, counterclockwise, or neither direction.

This randomization effectively counters existing keystroke prediction attacks by introducing variability in the key locations.
Every time users enter a character, the keyboard adopts one of two possible configurations, offset by one key width. This mechanism ensures that each character entry results in two possible configurations, assuming the user is not repeatedly entering the same character. Over the course of entering $n$ characters, the total complexity for predicting the entered content grows exponentially, reaching $2^{n}$.

Importantly, only the positions of the unselected keys can be affected by the randomization, and the shift by one key width does not alter the alphabetical order of the keys, nor does it lead to sudden and drastic changes in their spatial locations. Over time the keyboard may slowly rotate, and the keys will move to different positions, but the process is expected to be gradual and non-intrusive, and the color coding can assist users in quickly locating the approximate location of the keys when they are familiar with it. 
%Since users can seamlessly proceed with their input, the randomization is expected to be non-intrusive to their typing experience.

\subsection{Security Analysis}
The radial keyboard is robust against basic keystroke prediction methods because the keys are not in fixed locations.
It is also robust against brute-force attacks due to the exponential complexity. With $26\times29=754$ possible initial configurations (26 possible locations to insert the space and 29 possible rotations, i.e. 26 letters + space takes 2 + an extra for expanded key), the total complexity is $754 \times 2^{n}$ for entering $n$ characters. 
For $n>=80$, i.e., after typing about 10-20 words, the random space becomes larger than a fully randomized 3-row keyboard layout ($26! = 4.03\times10^{26}$)  

Since the randomization is introduced after each entry instead of intermittently, producing an exponential random space, and each keyboard configuration in this random space is independent and cannot be approximated by neighboring configurations, the uniform sampling strategy does not apply.
Furthermore, due to the space key being inserted in a random location among the letter keys, it is also impossible to assume a fixed relative transform between keys based on the alphabetical order, or separate the entry sequence into individual words and use short words like ``a'' or ``I'' to backtrack the configuration of the keyboard.
Although the relative transformations between keys are not entirely random, it remains computationally infeasible to incrementally predict the entered content. The first input yields 27 possible options (26 letters plus the space key), and each subsequent input expands the search space by at least a factor of 3. This is due to the possibility that the next key could be the previous key with a known offset, an adjacent key if a rotation occurred, or the space key. Even more possibilities exist considering in which direction the cursor moved from the previous key and whether the space key is included in computing the offset. Consequently, for an input sequence of length n, the number of potential paths grows to over $27 \times 3^{n-1}$.

\subsection{Potential Advanced Attack with Viterbi Decoding}
\label{sec:viterbi}
To assess the reliability of the radial keyboard, we also considered how an advanced attacker might exploit the consistent relative positioning of keys. A plausible strategy involves the use of Viterbi decoding with beam search guided by a probabilistic language model. In this approach, predictions are made iteratively, one character at a time. At each iteration, candidate sequences are ranked according to their likelihood based on a language model, such as a letter-level bigram or trigram, which estimates the probability of a letter appearing given one or more preceding letters. 
For example, in natural language, the letter ``b'' is more likely to be followed by ``u'' than by ``v''. To constrain the search space, only the top $n$ most probable candidates are retained at each step, where $n$ represents the beam size. This pruning strategy helps manage computational complexity while focusing on the most linguistically plausible predictions.
After considering all characters, the sequence corresponding to the maximum likelihood can be output as the final prediction.

\section{User Study}
\label{sec::study}

\subsection{Study Design}
To evaluate the security and usability of the radial keyboard, we conducted a user study with 3 conditions: the radial keyboard, ISPR~\cite{wan2024design}, and the standard QWERTY keyboard without any randomization. We adopted a within-subject design to minimize the influence of individual differences in typing speed and evaluation standard. See figure \ref{fig:teaser} for visual illustrations of the three methods.
As mentioned in section \ref{sec::attack}, ISPR represents the current state-of-the-art secure text-entry method with balanced security and usability performance, and the standard QWERTY keyboard is the commonly-used ``baseline'' condition without security protection.

We recruited 30 participants (17 female) between the ages of 18 and
34 (21 participants aged 25-34, and 9 aged 18-24)
from the university via flyers and mailing lists. The number of
participants was determined by G-Power with a medium effect size of
0.25 and a power of 0.8. There were no dropouts in the study. The
participants received a \$20 gift card after completing the study. The study was approved by the university's Institutional Review
Board (IRB).

\subsection{System and Apparatus}

We used a Meta Quest 3 tethered to a laptop with Nvidia GeForce RTX 4070, Intel i7-13700HX 2.1GHz, and 16GB RAM. The software was developed
using Unreal Engine 5.3.
For ISPR and QWERTY, the keyboard distance and scale were consistent with \cite{wan2024design}, which were reported in section \ref{sec::attack}.
The radial keyboard was positioned 1 meter in front of the user and the center of the keyboard was 0.25m below eye level. The radii for the inner keys and the entire keyboard were 7cm and 11cm, respectively. The scale, placement, and distance of the radial keyboard were determined based on feedback of the study team members who tested the system. Despite the difference in keyboard configurations, the visual size of the displayed and entered text were scaled to be consistent across the 3 conditions.

% \subsection{Procedure}
% \begin{figure}[]
%   \centering
%   \includegraphics[width=0.9\linewidth]{figures/flowchart.png}
%   \caption{\textcolor{red}{The flowchart of the study procedure. Each colored rectangle represents one trial or questionnaire component in the study procedure.}}
%   \label{fig:flowchart}
% \end{figure}

%Figure \ref{fig:flowchart} shows a flowchart of the study procedure. 
After reading the consent form and providing written consent, participants performed the text entry task using the 3 text entry methods in counterbalanced order. For each method, participants transcribed 3 phrases for practice, and another 10 for formal trials. They were instructed to type normally and prioritize the accuracy of transcription. The phrases to enter were displayed above the keyboard. Participants were introduced the mechanism of the method before they start typing, and for ISPR and radial conditions, they were informed that the randomization was for secure entry in case of confusion. 
The phrase sets we used were randomly drawn from the MacKenzie Phrase Set~\cite{mackenzie2003phrase} and counterbalanced across conditions, with manual adjustment to ensure that the word and character count for each set were roughly consistent. 
Afterwards, participants completed a questionnaire and participated in an semi-structured interview, during which they ranked the usability of the 3 methods and discussed the factors that influenced usability.

\subsection{Threat Model}

\paragraph{Keyboard Reconstruction}

Our threat model assumes that the attacker has access to the same application as the user and that the keyboard’s relative transform is consistent between them. This typically occurs when both rely on the default keyboard configuration without intentionally rescaling or adjusting the keyboard distance, or when the attacker has access to the modified configuration (e.g., by performing the attack on the user’s device).
Under this assumption, the attacker can measure the configuration of the keyboard, such as its size, distance, offset to floor or eye level, etc., through triangulation using multiple rays, and estimate the keyboard orientation using the mean HMD orientation during text entry.

In practice, the accuracy of these estimates depends on numerous factors, such as the choice of reference points and triangulation rays, as well as random hand jitter during measurement. However, because the goal of this study is to evaluate the security of text entry methods, we assume these measurements and estimates to be accurate rather than manually reconstructing the configuration. This avoids inflating the observed security performance due to incidental estimation noise. The results reflect a conservative lower-bound on security that is more informative than performance against a single, potentially noisy attack instance.

\paragraph{Keystroke Prediction}
%Since no keystroke prediction method has been developed specifically for the radial keyboard, and a comprehensive evaluation of adapted methods is beyond the scope of this project, we assumed an idealized scenario for the attacker in all conditions. 

For keystroke prediction, we assumed that the attacker has access to the controller's transformation data. This assumption is grounded in prior findings by \cite{wu2023privacy}, which demonstrate that sensor data remain unprotected in mainstream VR platforms. 
% Under such conditions, an adversary could feasibly replicate the user’s environment by downloading the same application and employing multiple rays to triangulate the keyboard’s transform.
% This assumption ensures a fair comparison across the three conditions and establishes a conservative lower bound for the security performance of each method. The reference values derived under this assumption remain relevant, even as prediction algorithms improve over time.
Specifically, the intersection of the ray with the keyboard plane is first computed to find the cursor location. Then, the keystroke is determined based on the cursor location, and the layout and mechanism of the corresponding keyboard.

\paragraph{Character Inference}

After predicting the keystrokes, we mapped the characters to the keys to uncover the entered text. For the standard QWERTY keyboard, the mapping is consistent and deterministic, but for ISPR and the radial keyboard, randomization is introduced to disrupt the mappings. Therefore, for both secure conditions we conducted both a basic attack and an advanced attack to thoroughly examine their robustness.

For ISPR, the basic attack assumed the ray to always start from the controller location, consistent with~\cite{wan2024design}. The advanced attack was the uniform sampling attack as described in section \ref{sec::attack}. 
To avoid bias in combining the 5 predictions in to a final prediction, we used GPT-4o with the following prompt: \textit{``You will be given 5 predictions of a phrase. Each prediction is only partially correct. Your goal is to predict what the original phrase is by combining the meaningful parts of the given predictions.''}. The predictions were also done in temporary mode with memory disabled to prevent GPT from reusing knowledge of prior predictions of the same sentence.

For the radial keyboard, the basic attack was based on a randomly generated initial location of the space key and the rotation of the keyboard, disregarding the rotation throughout the entry.
The advanced attack was the Viterbi decoding method discussed in section \ref{sec:viterbi}. We generated a trigram model using the \textit{big.txt} corpus from Norvig’s spelling corrector dataset~\cite{norvig2007spell}, which contains approximately one million commonly used English words. 
For the beam search, we used beam sizes of 200, 1000, 2000, and 4000. 
Notably, the exact expansion direction of a selected key depends on the direction from which the cursor enters, and the probability of post-entry shift of keys (left, right, or none) is also unknown to the attacker. Consequently, with the available information, it is not feasible to precisely simulate the keyboard's state transitions. To accommodate this uncertainty, we assumed an equal probability for each of the three possible shift directions (left, right, or none) following each key entry, which covers all possible outcomes.

\subsection{Measures}

We adopted the same measures as \cite{wan2024design} and \cite{wan2024analysis}, and evaluated the text entry methods in terms of security, typing performance, and usability. Although typing performance is closely tied to usability, we treat it separately from subjective usability measures in our analysis due to its objective nature.

\subsubsection{Security Evaluation}

Given the predictions and the corresponding ground truth, we computed the \textbf{Identical Character Ratio (ICR)}, which is the number of identical characters at the same position of the prediction and ground truth, divided by the total number of characters in the ground truth, and \textbf{Semantic Similarity (SS)}, which was computed as the cosine similarity of their semantic vector embeddings using the sentence transformer ``all-MiniLM-L6-v2''. Results for ISPR and the radial keyboard are reported separately for basic attacks (\textit{ISPR\_Basic, Radial\_Basic}) and advanced attacks (\textit{ISPR, Radial}).

\subsubsection{Entry Performance}

The entry performance is evaluated by entry rate in \textbf{words per
minute (WPM)}, and error rate per character entered, including \textbf{total error rate (TER)} and \textbf{not corrected error
rate (NCER)} during the formal trials of the text entry task.

%Typing performance was measured by entry rate and error rate. For entry rate we computed words per minute (WPM). For error rate we computed the corrected error rate (corrected error per character entered), and the total error rate consisting of the total number of corrected and uncorrected errors per character entered.

\subsubsection{Usability}
Usability was measured quantitatively using the \textbf{System Usability Scale (SUS)}~\cite{brooke1996sus} and the \textbf{raw NASA Task Load Index (TLX)}~\cite{hart2006nasa} questionnaires, and qualitatively through a semi-structured interview conducted at the end of the study with open-ended questions regarding the usability of the three methods.

% Not sure if we need a dedicated section for hypotheses

\subsection{Hypotheses}
We defined the following scientific hypotheses to evaluate the effects of radial keyboard.
\begin{itemize}
    \item \textbf{H1:} Participants would have lower ICR with the radial keyboard compared to ISPR and QWERTY keyboard.
    \item \textbf{H2:} Participants would have lower SS with the radial keyboard compared to ISPR and QWERTY keyboard.
    \item \textbf{H3:} Participants would type faster using QWERTY keyboard compared to ISPR and radial keyboard.
    \item \textbf{H4:} Participants would type more accurate using the radial keyboard compared to ISPR and QWERTY keyboard.
    \item \textbf{H5:} Participants would report better usability with the QWERTY keyboard compared with ISPR and radial keyboard.
    \item \textbf{H6:} Participants would report less task load with the QWERTY keyboard compared with ISPR and radial keyboard.
\end{itemize}

In terms of security, we hypothesized that the radial keyboard would offer significantly greater resistance to keystroke prediction attacks than the other two methods (H1 and H2) thanks to the more thorough randomization. For performance and usability, we hypothesized that the Radial keyboard would be more accurate than the other two methods due to the relaxed precision requirement (H4). We also hypothesized that the standard QWERTY keyboard would be more efficient and considered more usable than the two secure methods thanks to being more consistent (H3, H5, and H6), but the comparison between the radial and ISPR was exploratory.

% % TODO:: ISPR next to Radial for all figures
% \begin{figure}[]
%   \centering
%   \includegraphics[width=\columnwidth, trim = 0px 20px 0px 0px, clip]{figures/ICR_boxplot.png}
%   \caption[]{Box plots of identical character ratio (ICR).
%     The ICR results were significantly different between all conditions except for between ISPR\_Basic and Radial Keyboard.
%     Note: Throughout all figures, significance levels are indicated as follows: * $p<.05$, ** $p<.01$, *** $p<.001$.
%     }
%   \label{fig:icr}
% \end{figure}

% \begin{figure}[]
%   \centering
%   \includegraphics[width=\columnwidth, trim = 0px 20px 0px 0px, clip]{figures/SS_boxplot.png}
%   \caption[]{Box plots of semantic similarity (SS).
%     The SS results were significantly different between all conditions.}
%   \label{fig:ss}
% \end{figure}

% \begin{figure}[]
%   \centering
%   \includegraphics[width=\columnwidth, trim = 0px 20px 0px 0px, clip]{figures/TER_boxplot.png}
%   \caption[]{Box plots of total error rate results (TER).
%   There were no significant differences between conditions.
%     }
%   \label{fig:ter}
% \end{figure}

% \begin{figure}[]
%   \centering
%   \includegraphics[width=\columnwidth, trim = 0px 20px 0px 0px, clip]{figures/NCER_boxplot.png}
%   \caption[]{Box plots of not corrected error rate results (NCER).
%   There were no significant differences between conditions.
%     }
%   \label{fig:ncer}
% \end{figure}

\section{Results}
\label{sec::res}

\begin{figure*}[t]
  \centering
  \includegraphics[width=0.8\linewidth]{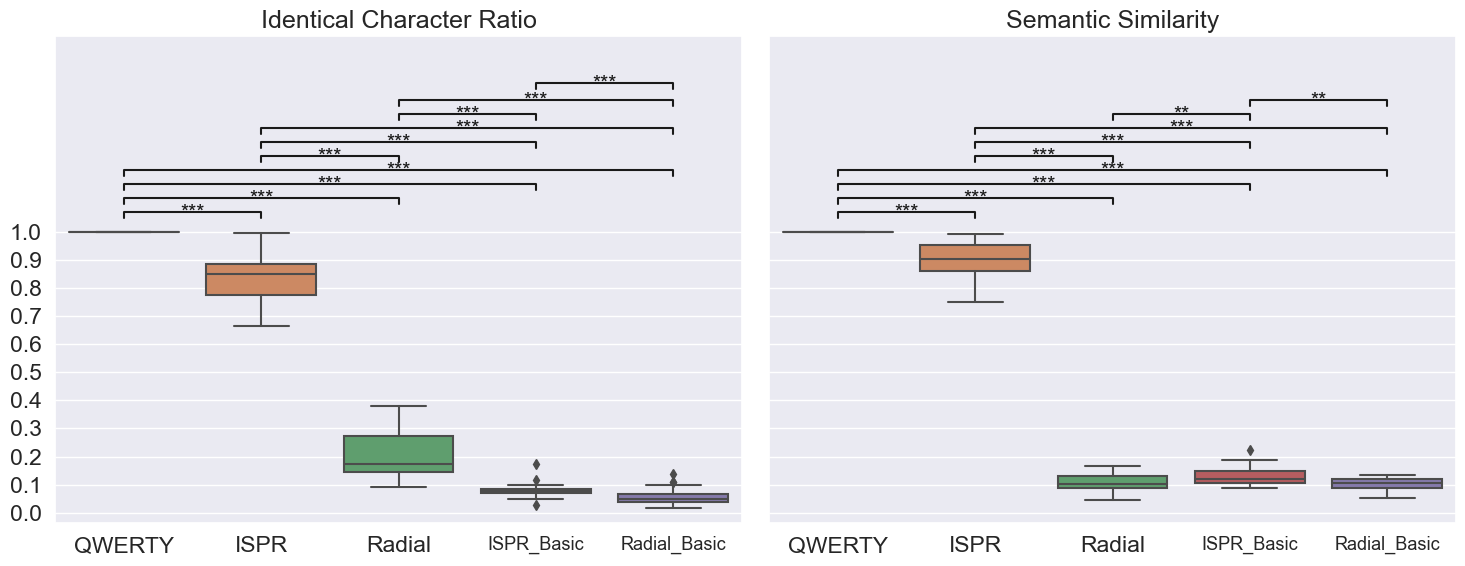}
  \caption[]{Box plots of identical character ratio (ICR) and semantic similarity (SS).
    The ICR and the SS results were significantly different between all conditions, except for the SS results between Radial and Radial\_Basic. 
    Note: Throughout all figures, significance levels are indicated as follows: * $p<.05$, ** $p<.01$, *** $p<.001$.
    }
  \label{fig:security}
\end{figure*}

\subsection{Quantitative Results}
We first conducted Shapiro-Wilk tests for all variables. 
% need update
All dependent variables showed significant violation of normality.
% For parametric data, we used repeated measures ANOVAs to analyze differences between the three conditions (QWERTY, ISPR, and radial keyboard) and reported descriptive statistics as mean (\textit{M}) and standard deviation (\textit{SD}).
For the non-parametric data, we used Friedman tests and reported descriptive statistics as median (\textit{Mdn}) and interquartile range (\textit{IQR}). 
Statistical tests assumed a significance value of $\alpha=.05$. 
When a Friedman test rejected the null hypothesis, we ran Conover’s all-pairs post-hoc test for Friedman-type (blocked) data with Holm correction for multiple comparisons (as implemented in JASP; R backend: PMCMRplus frdAllPairsConoverTest)..

\subsubsection{Security}
\label{sec:res-sec}

We observed that for both ICR and SS metrics, the prediction accuracy of the Viterbi attack plateaued at a beam size of 1000. Increasing the beam size further to 2000 and 4000 resulted in substantially longer computation times without any notable improvement in prediction performance. Therefore, for the Radial condition, we report results based on a beam size of 1000.

\paragraph{Identical Character Ratio}
Figure \ref{fig:security} shows the ICR result.
The analysis for ICR showed significant difference between five conditions $\chi^2(4)=115.71$, $p<.001$, $W=.96$.
Post hoc comparisons found the ICR in the QWERTY ($Mdn = 1$, $IQR=0$) condition was significantly higher than ISPR ($Mdn=.85$, $IQR=.11$),
$p<.001$, $r_{rb}=1$,
ISPR\_Basic ($Mdn = .078$, $IQR=.015$), $p<.001$, $r_{rb}=1$, Radial ($Mdn = .175$, $IQR=.127$), and Radial\_Basic ($Mdn = .049$, $IQR=.029$), $p<.001$, $r_{rb}=1$.
The ICR in ISPR condition was significantly higher than ISPR\_Basic, $p<.001$, $r_{rb}=1$, Radial\_Basic, $p<.001$, $r_{rb}=1$, and radial keyboard condition, $p<.001$, $r_{rb}=1$.
The ICR in Radial condition was also significantly higher than ISPR\_Basic, $p<.001$, $r_{rb}=1$, and Radial\_Basic, $p<.001$, $r_{rb}=1$.
Finally, ISPR\_Basic condition had significantly higher ICR than Radial\_Basic condition, $p<.001$, $r_{rb}=.6$.
%\textcolor{red}{No significant difference was found between ISPR\_Basic and radial keyboard condition, $p=.07$}.
These result support hypothesis H1.

\paragraph{Semantic Similarity} 
Figure \ref{fig:security} shows the SS result.
The analysis for SS showed significant difference between five conditions $\chi^2(4)=98.43$, $p<.001$, $W=.82$.
Post hoc comparisons found the SS in the QWERTY ($Mdn = 1$, $IQR=5.4\times10^{-8}$) condition was significantly higher than ISPR ($Mdn=.90$, $IQR=.091$), $p<.001$, $r_{rb}=1$, ISPR\_Basic ($Mdn = .12$, $IQR=.044$), $p<.001$, $r_{rb}=1$, Radial ($Mdn = .10$, $IQR=.04$), $p<.001$, $r_{rb}=1$, and Radial\_Basic ($Mdn = .11$, $IQR=.03$), $p<.001$, $r_{rb}=1$.
The SS in ISPR condition was significantly higher than Radial, $p<.001$, $r_{rb}=1$, Radial\_Basic, $p<.001$, $r_{rb}=1$, and ISPR\_Basic, $p<.001$, $r_{rb}=1$.
Finally, ISPR\_Basic condition had significantly higher SS than Radial\_Basic condition, $p=.005$, $r_{rb}=.59$, and Radial condition, $p=.006$, $r_{rb}=.49$.
These result support hypothesis H2.

\subsubsection{Text Entry}
\begin{figure}[t]
  \centering
  \includegraphics[width=\columnwidth]{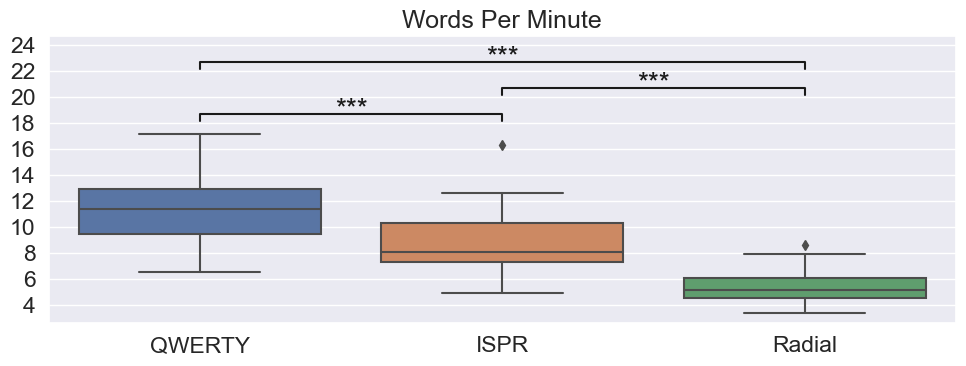}
  \caption[]{Box plots of words per minute (WPM).
    The WPM results were significantly different between all conditions.}
  \label{fig:wpm}
\end{figure}

\paragraph{Words Per Minute}
Figure \ref{fig:wpm} shows the WPM result.
The analysis for WPM showed significant difference between three conditions $\chi^2(2)=49.40$, $p<.001$, $W=.82$.
Post hoc comparisons found participants in the QWERTY condition ($Mdn=11.39$, $IQR=3.46$) typed significantly faster compared 
with ISPR condition ($Mdn=8.03$, $IQR=2.98$), $p<.001$, $r_{rb}=.86$,
and the radial keyboard condition ($Mdn=5.13$, $IQR=1.56$), $p<.001$,, $r_{rb}=1$
ISPR condition was also significantly faster than the radial keyboard condition, $p<.001$, $r_{rb}=1$,
These results support hypothesis H3.

\begin{figure}[t]
  \centering
  \includegraphics[width=\columnwidth]{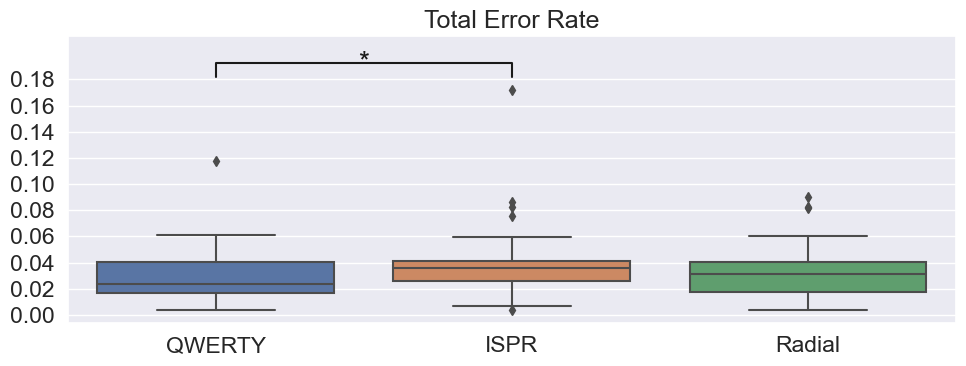}
  \caption[]{Box plots of Total Error Rate Per Character (TER). The TER was significantly higher in the ISPR condition compared to the QWERTY condition.}
  \label{fig:ter}
\end{figure}

\paragraph{Total Error Rate}
Figure \ref{fig:ter} shows the TER result.
Analysis results of the TER indicated significant difference between three conditions $\chi^2(2)=6.07$, $p=.048$, $W=.10$.
Post hoc comparisons found participants in the QWERTY ($Mdn=.024$, $IQR=.023$) typed significantly more accurate than the ISPR condition ($Mdn=.036$, $IQR=.015$), $p=.04$, $r_{rb}=.5$.
No significant difference was found between the radial keyboard condition ($Mdn=.031$, $IQR=.023$) and the other two conditions.
This result does not support H4. 

\paragraph{Not Corrected Error Rate}
Analysis results of the NCER indicated no significant difference among the 
QWERTY ($Mdn=0$, $IQR=.004$), 
ISPR ($Mdn=0$, $IQR=.003$) and 
radial keyboard conditions ($Mdn=.004$, $IQR=.007$), $\chi^2(2)=3.85$, $p=.15$. 
This result also does not support H4.

\subsubsection{Usability}
\begin{figure}[]
  \centering
  \includegraphics[width=\columnwidth]{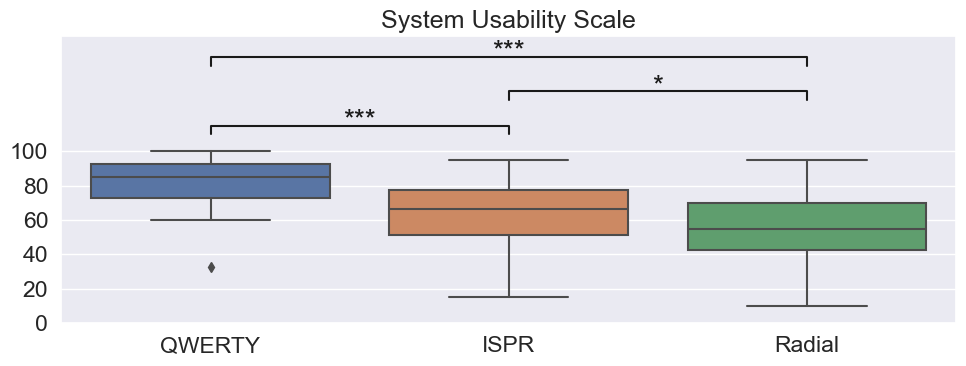}
  \caption[]{Box plots of System Usability Scale (SUS).
  The SUS results were significantly different between all conditions.
    }
  \label{fig:sus}
\end{figure}

\paragraph{System Usability Scale}
Figure \ref{fig:sus} shows the SUS result.
The analysis for SUS scores showed significant difference between three conditions $\chi^2(2)=32.43$, $p<.001$, $W=.54$.
Post hoc comparisons found the usability of the QWERTY condition ($Mdn = 85$, $IQR = 19.375$) was significantly higher than the 
ISPR ($Mdn = 66.25$, $IQR = 26.25$), $p<.001$, $r_{rb}=.86$ and radial keyboard conditions ($Mdn = 55$, $IQR = 27.5$), $p<.001$, $r_{rb}=.97$.
ISPR condition's SUS scores were also significantly different from the radial keyboard condition, $p=.02$, $r_{rb}=.46$.
These results support hypothesis H5.

% \paragraph{SUS Subscales}
% We only
% SUS 1

\begin{figure}[t]
  \centering
  \includegraphics[width=\columnwidth]{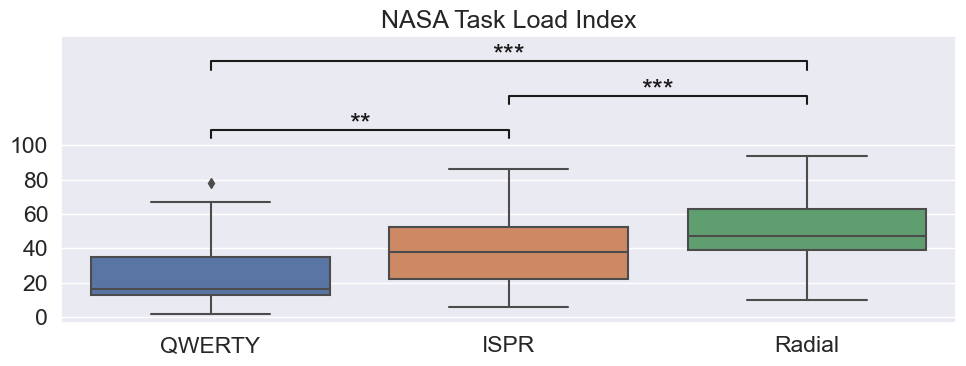}
  \caption[]{Box plots of NASA Task Load Index (TLX).
  The TLX results were significantly different between all conditions.
    }
  \label{fig:tlx}
\end{figure}

\paragraph{NASA Task Load Index}
Figure \ref{fig:tlx} shows the NASA TLX result.
The analysis for NASA TLX scores showed significant difference between three conditions $\chi^2(2)=28.07$, $p<.001$, $W=.47$.
Post hoc comparisons found the task load of the QWERTY condition ($Mdn = 16.5$, $IQR = 22$) was significantly lower than the 
ISPR ($Mdn = 38$, $IQR = 30$), $p<.001$, $r_{rb}=-.71$ and radial keyboard conditions ($Mdn = 47$, $IQR = 23.5$), $p<.001$, $r_{rb}=-.98$.
ISPR condition's NASA TLX scores were also significantly lower than the radial keyboard condition, $p=.002$, $r_{rb}=-.51$.
These results support hypothesis H6.

\begin{figure}[t]
  \centering
  \includegraphics[width=\columnwidth]{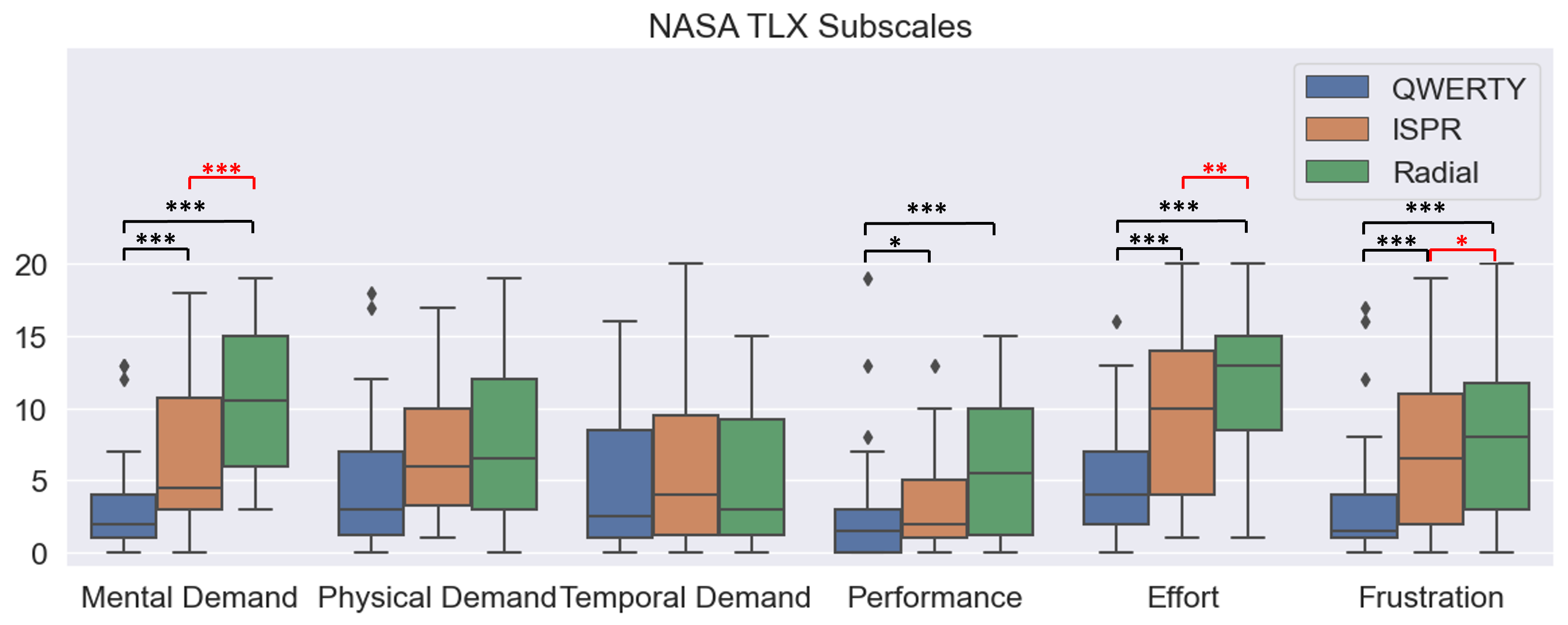}
  \caption[]{Box plots of NASA TLX subscales. Participants reported higher mental demand, effort, and frustration in the radial keyboard condition compared to the ISPR condition. Significant differences between the ISPR and the radial keyboard conditions are highlighted in red.
    }
  \label{fig:tlx_sub}
\end{figure}

\paragraph{NASA TLX Subscales}
Figure \ref{fig:tlx_sub} shows more detailed comparisons of the TLX subscales.
We only reported cases where ISPR is significantly different from the radial keyboard as we are most interested in the difference between the two secure keyboard designs.
The analysis for NASA TLX mental demand subscale showed significant difference between three conditions $\chi^2(2)=31.22$, $p<.001$, $W=.52$.
Post hoc comparisons found the mental demand of the ISPR condition ($Mdn = 4.5$, $IQR = 7.75$) was significantly lower than the radial keyboard condition ($Mdn = 10.5$, $IQR = 9$), $p<.001$, $r_{rb}=-.66$.
The analysis for NASA TLX effort subscale showed significant difference between three conditions $\chi^2(2)=30.78$, $p<.001$, $W=.51$.
Post hoc comparisons found the effort of the ISPR condition ($Mdn = 10$, $IQR = 10$) was significantly lower than the radial keyboard condition ($Mdn = 13$, $IQR = 6.5$), $p=.004$, $r_{rb}=-.47$.
The analysis for NASA TLX frustration showed significant difference between three conditions $\chi^2(2)=24.24$, $p<.001$, $W=.40$.
Post hoc comparisons found the frustration of the ISPR condition ($Mdn = 6.5$, $IQR = 9$) was significantly lower than the radial keyboard conditions ($Mdn = 8$, $IQR = 8.75$), $p=.03$, $r_{rb}=-.36$.
%These results also support hypothesis H6.

\subsubsection{Learning Effects}
We are also interested in examining whether participants' typing speed showed significant learning effect over the course of the task. 
We conducted Mann-Kendall trend tests on characters per minute (CPM) across phrases for each participant in each input condition.
The results indicated significant increasing trend ($\tau > 0,p<.05$) in 7 participants (23.3\%), decreasing trend ($\tau < 0,p<.05$) in 1 participant, and no significant monotonic trend ($p>=.05$) in 22 participants for the radial keyboard condition.
For QWERTY and ISPR, significant increasing trend were only observed in 1 (3.3\%) and 2 (6.7\%) participants, respectively. The rest were associated with no significant trend.

\subsection{Qualitative Results}

% https://miro.com/app/board/uXjVI3NVEzM=/
% https://docs.google.com/spreadsheets/d/13evlQ0kmsqkZIa57cU9QItwmsDkebjklYp7BhZGJiVE/edit?gid=438415321#gid=438415321
% \subsubsection{Physical demanding manifests differently in ISPR and Radial}

We conducted a thematic analysis~\cite{braun2006using} to analyze qualitative data. The first author initiated the open coding phase, after which the first two authors collaboratively reviewed and organized the codes using constant comparison, constructing an affinity diagram in the process. Each code was examined in relation to existing categories to determine its most appropriate placement. Discrepancies were discussed and resolved collaboratively, with input from a third author when needed. As higher-order themes began to emerge, the full research team participated in iterative discussions to assess their relevance and contribution to our research aims. 

\subsubsection{Radial keyboard reduces physical demand}
\label{sec:qualitative 1}
%Participants experienced physical demands differently across input methods. %Radial was generally less taxing due to its compact, centralized layout, while ISPR imposed strain through unpredictable shifts in the ray origin (SP), requiring frequent recalibration of movement.
%
Although, users might engage in unnecessary movements when using the radial keyboard due to unfamiliarity, the radial layout was often praised for minimizing arm movement by placing all keys within a spatially compact and centralized structure, allowing each key to be equally accessible and the hand to remain largely stationary during entry. As one participant described, ``Equal access to all of the keys… it’s a circle, you have equal access to all of the keys… whereas, like on the keyboard, you’re like moving your one hand around.'' [P7] Another emphasized, ``I like how they’re close together so that my hand can just like stay here, and it’s the least physically demanding.'' [P13] Even when input speed suffered, participants found radial keyboard to be physically manageable: ``I actually moved slower, but it wasn’t as physically taxing.'' [P16]
%However, for Radial, due to the lack of familiarity, users might engage in unnecessary repositioning. One participant noted, ``It felt like it was a lot of movement… pretty small areas to pick… like you needed too much.'' [P30]

ISPR’s physical demand arose from its randomization mechanism: the ray’s starting point (SP) changes after random intervals. This dynamic required users to adapt constantly to differing mappings between wrist movement and cursor movement. When SP moved far forward, users had to rotate their wrist to more extreme angles to get to the targets; when SP shifted behind, precise input required the wrist to be extra steady. One participant observed, ``ISPR is hard to calibrate when SP is far ahead; behind is easier due to less movement required.'' [P6] Another echoed, ``They’re jumping around, makes it physically difficult to get the spot where I need to.'' [P17]
These spatial recalibrations often led to physical fatigue and discomfort: ``It's more physically demanding (ISPR) because my arms were hurting. I was also making a lot of mistakes.'' [P16] In some cases, the sudden change in SP also produced a sensory mismatch that participants found jarring or unpleasant: ``When it suddenly changes to like behind me… it feels almost like, ticklish… I just really hate that feeling… Sometimes the ray is gonna cut through you.'' [P13]

\subsubsection{Radial keyboard elevates cognitive demand through unfamiliar visual search}
\label{sec:familiarity}

Participants reported experiencing a higher cognitive demand when using the radial keyboard, largely due to unfamiliarity with its visual layout and the spatial inconsistencies caused by rotation. In contrast to the QWERTY layout, where participants could rely on well-developed spatial and muscle memory, the radial keyboard required active visual search, especially during early use. This cognitive demand was further amplified by the randomized key shifts, causing the absolute position of each key to change over time.
For many, this unfamiliarity created cognitive overhead. One participant remarked: ``I'm less familiar (with Radial). I mean, I know the alphabet, but... when it would rotate especially, it would be hard to remember where a certain key is... the normal QWERTY one, I've been typing on that for probably 20 years.'' [P8] Another described the experience as disorienting: ``I didn't know what exactly was the different keys located... I had to just go round and round... I couldn't catch a pattern.'' [P14]

Participants often compared the radial layout with QWERTY, noting that the latter allowed automatic recall of key positions, whereas the former required conscious effort: ``I still had to think a little bit of like, where's that? Where’s that gonna be?'' [P23] and ``I'm just not really used to... typing and thinking... in alphabetical order.'' [P21]
Encouragingly, some participants anticipated that familiarity would improve with continued exposure. As one expressed: ``QWERTY took me forever to learn... but for radial, maybe an hour to an hour and a half, I feel like I'm confident doing it.'' [P21] Visual aids like color coding offered some facilitation. While not uniformly effective in a short session, participants noted its potential: ``I think it helps that there were different colors and sometimes you kind of remember what area it should be in because of the colors.'' [P3] and ``Maybe after using it for a long time... you’d develop the sense that `A' is red, `I' is green... so you kind of know which way to go.'' [P23]

% These findings suggest that Radial’s cognitive demand originates not from key content (e.g., alphabetical order), but from the disruption of spatial predictability and visual search heuristics—barriers that could be reduced through layout stabilization and reinforced visual cues.

% Visual aids, such as color coding, were mentioned as mitigating factors, though their efficacy was inconsistent across users: ``The color might help remember given enough practice'' [P8], while another admitted, ``Didn’t pay attention to the colors'' [P11].

\subsubsection{Sources of error vary across input methods}
\label{sec:error}

%- Radial extra space reduces the physical demand
%- Radial key color coding  reduces the cognitive demand

Participants reported that the type and frequency of errors differed across the three input methods.
With ISPR, a dominant source of error came from the unpredictable jumps of the ray’s SP. Each jump can cause a different key to be selected than intended, leading to errors especially when typing quickly. As one participant described, ``When you get used to the ray and then, and the second moment it changed'' [P1] Another added, ``I think, is quite annoying, because, um, sometimes I just choose a character, but actually it is jumping to another one.'' [P9]
Another source of error shared across ISPR and QWERTY was hand instability. Despite QWERTY’s static ray origin, precision remained challenging for some participants: ``QWERTY sometimes miss the keys.'' [P6], and as mentioned in section \ref{sec:qualitative 1}, ISPR required even more stability when SP is behind the user.

With a radial keyboard, such errors were less reported because of the dynamic resize of the keys and infinite selection area. As some participants noted, ``It was very helpful having them enlarged because I felt more confident pressing it.'' [P23] and `` I really liked the idea of I don't have to like hit the button.'' [P21].
However, several participants noted that double-clicking the same key often led to mis-entry. 
One participant explained: ``You press `F' two times... and you don't go back to the center. So when you're pressing it two times, it changes... you're either landing at `E' or a `G'.'' [P5]
%This problem was particularly prominent when quick repetition combined with minor cursor drift caused by hand instability.
%To mitigate this issue, visual enhancements such as key enlargement and color coding, helped some users reduce errors and improve confidence. Others echoed that enlarged targets offered reassurance, especially when near critical keys like "Enter". 

\section{Discussion}

\subsection{Security}

The results of ICR and SS in section \ref{sec:res-sec} indicate that the radial keyboard provide strong security, as both basic and advanced keystroke prediction attacks targeting the radial keyboard with precise cursor information were largely ineffective in recovering either the character sequence or the semantic content of the input.
ISPR demonstrated comparable security to the radial keyboard when attacked with a basic method assuming a constant ray starting point, which is consistent with the findings in \cite{wan2024design}. However, with the uniform sampling attack of five samples, most entered phrases could be reconstructed verbatim, and the accuracy might improve further with more samples or a custom-trained language model. This exposes a critical security vulnerability that was overlooked in the original paper, and indicates that intermittent randomization in a limited range, despite effective against basic attacks, may be susceptible to more sophisticated prediction attacks. In contrast, these limitations were effectively mitigated by the radial keyboard’s more thorough per-keystroke randomization, which allows any key to appear in any sector. The results suggest that fully decoupling a key’s position from its identity significantly enhances resistance to prediction attacks. Notably, the radial layout enables this level of security without disrupting the relative arrangement of the keys, which is more advantageous than the more common grid-based layout.

It is important to note that, although the user study demonstrated strong security performance of the radial keyboard against our implementation of the Viterbi attack, we cannot rule out the possibility that alternative attack strategies may yield improved prediction performance. For example, attackers could employ language models trained on different corpora, or adopt entirely different decoding methods. While a comprehensive exploration or optimization of attack strategies is beyond the scope of this paper, we encourage future work to examine the robustness of the radial keyboard under a broader range of threat models.

\subsection{Entry Performance and Usability}

The entry performance result shows that typing with the radial keyboard was the slowest among all three methods. The survey and interview responses provide insights to the performance differences and suggest improvement over time.

% Supplement with qualitative results
% TLX
According to the NASA TLX and the SUS results, the radial keyboard was associated with significantly higher task load and lower perceived usability than ISPR and QWERTY, particularly due to higher mental demand, effort, and frustration. This could be explained by the participants being less familiar with alphabetical layout compared to the QWERTY layout, and the rotation that moved the keys around required them to utilize the relative position and/or the colors of the keys, which was different from the absolute position they typically relied on when typing on a QWERTY keyboard. Both factors led to additional mental demand and effort to search for the keys and made typing slower, and the slower-than-desired entry led to elevated frustration.

The impact of familiarity is also supported by participants' qualitative responses, particularly highlighted in section \ref{sec:familiarity}. All participants mentioned their familiarity with the layout as the main factor that made typing easy in QWERTY or ISPR, and the majority of them considered unfamiliarity to be the primary factor limiting their performance when typing with the radial keyboard.
Most participants expected improved performance after more practice. Although some raised concerns about the inconsistent positions of the keys making the learning process difficult, others acknowledged that the order of the keys were consistent regardless of where they were rotated to, so they could always find the next key based on the relative position, e.g., ``a'' and ``b'' are next to each other, whereas ``o'' is across, and some pointed out that the color coding of the keyboard were helpful for quickly locating the target letters. In fact, the impact of keyboard rotation on usability can be reduced by making the rotation probability equal or near equal for both directions. This reduces the chance for the keys to keep moving in one direction and end up drastically away from their original position for each entry session, without reducing the exponential random space. Considering that the initial rotation of the keyboard as well as the location of the space being randomized for each entry session, the probability of a key to appear in any location are still equal across sessions, so it is still impossible to predict a key based on its absolute position.

% SUS (No overall significance, consider omitting)
%Regarding the SUS, the standard QWERTY was rated significantly more usable than the two secure entry methods, but no significant difference was found between the radial keyboard and ISPR for the total scores. The sub-scales that did show significant differences were \textit{``I think that I would like to use this system frequently''}, \textit{``I thought the system was easy to use''}, and \textit{``I felt very confident using the system''}, where the scores for ISPR were higher. 
%These results make sense considering the higher task load and slower entry rate when typing with the radial keyboard, and participants were more confident typing with the QWERTY layout which they were more familiar with.

The only notable usability issue raised by participants was the double-clicking problem discussed in section~\ref{sec:error}. This issue was a significant source of error for the radial keyboard and largely offset the accuracy advantage we had anticipated. Although we ensured that the random key shift would not change the key under selection if the controller remained stationary, mis-entry could still happen since participants were mostly aiming at the center of the key when entering, which would turn into the boundary if the key shift happened, so a slight hand jitter could lead to the selection of the adjacent key instead. This issue can be resolved by increasing the size of the selected key to be three times of the width of an unselected key instead of twice, so that the randomization after each key press could result in three different outcomes. If we reject the outcome that would bring the boundary of the selected key closest to the cursor position and always choose from the other two instead, a slight hand jitter should not lead to mis-selection of adjacent keys, and the random space would remain unaffected.

We also noticed that when typing with the radial keyboard, a few participants would still try to carefully aim directly at the small keys with the ray, which was unnecessary and made the entry slower and less accurate. One participant explained that they were more used to pointing directly at the target than pointing outside when selecting. Although this might not be an issue for everyone, a better visual indicator to convey that the selection area extends outside of the keys could be helpful and easy to implement, such as boundary lines that extend into infinity, or the keys fading as they go outwards.

% Conclusion on Usability
Overall, the results suggest that the relatively higher difficulty of using the radial keyboard was largely due to unfamiliarity with a system being used for the first time, rather than a fundamental problem with the radial keyboard design.
Although the short practice time is a key limitation of this study, both the quantitative and qualitative results suggest improvement over time.
Furthermore, as highlighted in section \ref{sec:qualitative 1}, some participants appreciated the ease of selection using the radial keyboard with the virtual key expansion, relaxed precision requirement, and equal distances to access all keys. These usability features are difficult or even impossible to implement on a QWERTY layout, which leads to the discussion of whether the dominance of QWERTY keyboards is a limiting factor in the design and adoption of novel text entry methods that could potentially achieve better usability and performance. 

% Discussions of long-term improvement
\subsection{Keyboard Layout for VR: QWERTY and Beyond}

% QWERTY is the standard only because of familiarity
The QWERTY layout was technically designed to prevent jamming when typing on typewriters~\cite{yasuoka2011prehistory}, a reason that has long been irrelevant, by separating letters that frequently appear together farther apart. 
Typing in VR presents a fundamentally different interaction paradigm compared to traditional interfaces, particularly when using a virtual keyboard. Users must explicitly aim at targets, and many prefer to type with only their dominant hand due to the difficulty of precise aiming with the non-dominant hand. Under these conditions, typing on a QWERTY keyboard in VR results in a high index of difficulty according to Fitts' Law~\cite{fitts1954information}, due to the small target widths and the large movement distances required between consecutive key selections. This significantly compromises ergonomics in a way that is not easily mitigated through practice.
In contrast, the radial keyboard offers a much larger effective key width, as keys extend infinitely outward, making selection easier and less physically demanding. This advantage was also recognized by several participants, as noted earlier. Additionally, unlike the QWERTY layout, the alphabetical arrangement does not deliberately separate frequently co-occurring letters.
In fact, some participants suggested that the radial keyboard could be further improved by grouping frequently used letters together, which they felt might outperform the current alphabetical layout in the long run.

% Familiarity is not a permanent issue
While most people understand alphabetical order, they are not as familiar with typing using this layout compared to QWERTY, which they have been doing since they started learning how to type. However, familiarity improves over time with practice. Considering that people are able to learn the QWERTY keyboard, despite it having a somewhat arbitrary key layout, and still consider it to be highly usable after sufficient practice, it is reasonable to expect that they will be able to adapt over time to alternative layouts that use a more intuitive key ordering.

For this exploratory study, we do not expect a novel method with roughly a minute of practice to outperform the QWERTY keyboard in terms of entry efficiency.  A longitudinal study would be necessary to formally assess the usability of the radial keyboard after more extensive use.  However, based on the results from this study, the radial keyboard is a promising secure text entry method for VR. It offers stronger security and several usability advantages over existing text entry methods, and familiarity issues may be mitigated with practice. These together provide a good reason to consider the radial keyboard for applications where security is important.

\section{Conclusion and Future Work}
In this study, we presented the uniform sampling attack to illustrate the vulnerability of the current state-of-the-art secure text entry methods based on intermittent randomization within a limited range. We addressed the limitations of existing methods by proposing a novel virtual radial keyboard that introduces full randomization to the absolute locations of the keys while keeping the relative order consistent. The radial keyboard also supports dynamic resize of virtual keys and selection areas that extends into infinity to facilitate the ease of key selection.
The user study demonstrated that the radial keyboard significantly outperformed the state-of-the-art in terms of security, but with lower entry rate and increased user-reported task load which were largely due to unfamiliarity with a non-QWERTY layout. While trade-offs are expected between security and usability, both the quantitative and qualitative results suggest improvement over time, and the usability features that come with the radial keyboard were appreciated by some participants over the QWERTY keyboard.
Overall, the results suggest that the radial keyboard is a promising approach for situations where secure entry is necessary in VR, and we encourage future studies to further explore the potential of this idea by evaluating the performance of experienced users and considering possible user interface improvements, such as alternative layouts and key visualizations. 

\bibliographystyle{abbrv-doi}

\bibliography{reference}
\end{document}